\documentclass{PoS}

\usepackage{amsmath,amssymb}
\usepackage{graphicx}
\usepackage{verbatim}
\usepackage{color}
\usepackage{soul,xcolor}

\newcommand{\be}{\begin{equation}}
\newcommand{\ee}{\end{equation}}
\newcommand{\ba}{\begin{eqnarray}}
\newcommand{\ea}{\end{eqnarray}}
\newcommand{\ban}{\begin{eqnarray*}}
\newcommand{\ean}{\end{eqnarray*}}

\title{Consequences of conformal anomaly on fluid dynamics}

\ShortTitle{Consequences of conformal anomaly for fluid dynamics}

\author{\speaker{Alina Czajka}\\
        National Centre for Nuclear Research\\
        E-mail: \email{alina.czajka@ncbj.gov.pl}}

\abstract{We review a recent progress on fluid dynamics applied to strongly interacting nuclear matter. The efforts are made to highlight consequences of scale invariance breaking on the hydrodynamic description of a nuclear medium. Both phenomenological and analytical findings are summarized.}

\FullConference{XIII Quark Confinement and the Hadron Spectrum - Confinement2018\\
		31 July - 6 August 2018\\
		Maynooth University, Ireland}

\begin{document}

\section{Introduction}

Relativistic fluid dynamics provides an effective framework to investigate thermodynamic and transport properties of strongly interacting matter produced at the Relativistic Heavy-Ion Collider (RHIC) and the Large Hadron Collider (LHC). The dynamics of nuclear matter produced in nuclear collisions is governed by the quantum chromodynamics (QCD), but due to the complexity of the theory and lack of systematic methods to study the system from first principles combining data-driven observations and fluid dynamic equations provides a very efficient approach to explain an evolution of the system. The application of relativistic fluid dynamics to model the physics of heavy ion collisions is reviewed in, for example, \cite{Gale:2013da,Heinz:2013th}. Similarly, the development in formulation of modern hydrodynamic frameworks is discussed, for example, in \cite{Florkowski:2017olj,Romatschke:2017ejr}. The successful employment of fluid dynamics to study collective phenomena of the hadronic matter motivates well its further development and testing its limits. 

First applications of fluid dynamics to model the behavior of nuclear matter assumed that the system is conformally invariant and the specific shear viscosity must be small. Later on an enormous progress has been made; several formulations of viscous hydrodynamics have been developed and conformal anomaly consequences have been elaborated. The effects of the conformal symmetry breaking are significant in multiple ways for a fluid dynamics description. In particular, they determine a form of bulk viscous correction to the phase space density, which in turn is meaningful for an implementation of the Cooper-Frye prescription or studying electromagnetic probes in heavy-ion collisions. What is more, the bulk viscous correction enables one to compute transport coefficients emerging when the scale symmetry is broken. The bulk viscosity coefficient and its constraints can be then studied.

In these proceedings, we review the recent progress on phenomenological and analytical findings coming from the application of fluid dynamics to model and characterize the properties of strongly interacting matter. We focus on the physics determined by the conformal symmetry breaking.

\section{Impact of conformal anomaly on phenomenological results}

An increasing interest in the implications of scale symmetry breaking into hydrodynamic modeling of heavy-ion collisions has been stimulated by the lattice QCD results revealing an enhanced value of the trace anomaly around the critical temperature. Since the trace anomaly is expected to govern the behavior of bulk viscosity, the lattice findings induced many attempts to determine the spectral function of bulk viscosity. Such an attempt was undertaken in~\cite{Karsch:2007jc}, where the temperature dependence of bulk viscosity was obtained. For this purpose, an ansatz for the low-frequency spectral function was proposed and then lattice QCD data for the trace anomaly were used. Although the final height and shape of the bulk viscosity as the function of temperature requires more studies \cite{Moore:2008ws,Romatschke:2009ng}, it is rather expected that since that bulk viscosity is dictated by the trace anomaly, it has a peak around the critical temperature. Such a guidance inspired various models to explore how the violation of conformal symmetry influences the hydrodynamic description of strongly interacting matter evolution.

\begin{figure}[!t]
	\begin{center}
		\includegraphics[width=0.6\textwidth]{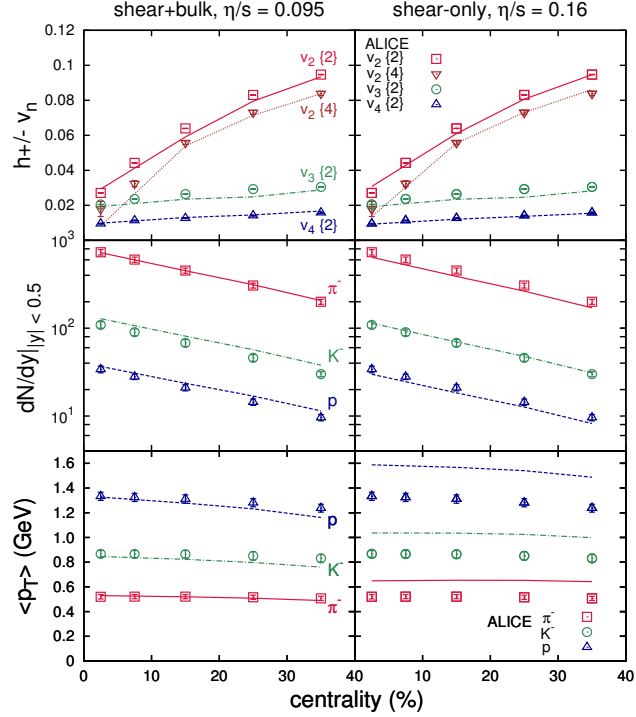}
		\caption{Integrated $v_n$ (upper),
			mid-rapidity multiplicity $dN/dy|_{y=0}$ (middle)
			and mean-$p_T$ (lower) are shown as functions of centrality.
			The experimental data are from the ALICE experiment: $v_n$ from \cite{ALICE:2011ab} and $dN/dy|_{|y|<0.5}$ and $\langle p_T \rangle$ from \cite{Abelev:2013vea}. The model predictions in the left column take into account the bulk viscosity while those from the right one do not. The figure is taken from Ref.~\cite{Ryu:2017qzn}.
		}
		\label{fig1}
	\end{center}
\end{figure}

In Ref.~\cite{NoronhaHostler:2008ju} the bulk viscosity at low temperature was studied within hadron resonance gas model. It was shown that the bulk viscosity increases when the temperature approaches the critical value. Later, it was tested how physical observables, such as $\nu_2$, change when bulk viscosity, obtained with the lattice equation of state, is included \cite{Denicol:2009am,Monnai:2009ad,Song:2009rh}. Recently, the impact of strongly peaked bulk viscosity at the phase transition region, with the value $\zeta/s \approx 0.3$ at the peak, on different experimentally accessible quantities was elaborated in several works. Integrated $\nu_n$, mid-rapidity multiplicity and mean transverse momentum were studied in \cite{Ryu:2015vwa,Ryu:2017qzn}. The comparison of these observables with and without incorporation of bulk viscosity is presented in Fig.~\ref{fig1}. The main finding is that the whole set of the hadronic observables can be consistently described when bulk viscosity is included. With the same set-up of the parameters defining bulk viscosity the direct photon spectrum and $\nu_2$ were investigated in Ref.~\cite{Paquet:2015lta}. It was shown that the effect of bulk viscosity on the photon spectrum is small and it causes a slight softening of it. However, $\nu_2$ as a function of mean transverse momentum changes its shape and contributes to the shift of its maximum when bulk viscous effects are included. Later, the interferometry correlations were explored in \cite{Bozek:2017kxo}. It is argued that the strongly peaked bulk viscosity leads to a slight increase in the life-time of the system produced by central Pb+Pb collisions (at 2760 GeV), which is related to slowing down of the outer layers of the fireball. This phenomenon has an impact on the interferometry radii making a small reduction in the ratio $R_{out}/R_{side}$, which improves the agreement with experimental data.

Other phenomenological consequences of enhanced bulk viscosity near the critical point were elaborated in Ref.~\cite{Monnai:2016kud}, where the form of the ratio of bulk viscosity to shear viscosity is motivated by the result found via holographic techniques, that is, it linearly depends on the conformality breaking. It was pointed out, in particular that the critically enhanced bulk viscosity gives rise to an increase in particle spectra at forward-midrapidity and this information may be significant for searches of the critical point of the QCD. 

Data-driven findings provide many tests and constraints on transport properties of strongly interacting matter confirming importance of the bulk viscosity importance for hydrodynamic simulations. These, however, need strong support from analytical approaches, whose development is shortly discussed below.

\section{Nonconformal fluid dynamics with mean field effects}

There are two ingredients of the fluid dynamics which play a key role when the scale symmetry is broken in a system under consideration. These include a viscous correction to the phase space density and a bulk viscosity coefficient together with its relaxation time. Their dependence on possible conformal symmetry breaking parameters can be consistently established in the limit where systematic analytical methods are applicable. Within such a ramification any system whose characteristics undergo a perturbative expansion can be considered. A convenient framework is provided by a kinetic theory applicable in high enough energy regime and, in particular, when its quasiparticle description holds. Importantly, such a description must be considered when nonconformal fluid dynamics is explored as one has to include all effects responsible for the violation of the scale symmetry, also temperature dependent ones. In fact, the parameters that lead to the conformal anomaly are the zero-temperature mass of the system's constituents $m_0$ and the position-dependent thermal mass $m_{\rm eq}(x)$, which results in the emergence of the Callan-Symanzik $\beta_\lambda$ function determining the running of the coupling constant. The concepts of such an effective kinetic theory allows one to formulate equations of viscous nonconformal fluid dynamics.

Many attempts were undertaken to provide a consistent description of the nonconformal fluid evolution, see \cite{Sasaki:2008fg,Chakraborty:2010fr,Bluhm:2010qf,Dusling:2011fd,Romatschke:2011qp,Albright:2015fpa,Chakraborty:2016ttq,Tinti:2016bav}. They have provided a significant progress in understanding various medium effects and, in particular, allowed one for an extraction of the bulk viscous correction $\Delta  f$ to the phase space density being linearly dependent on the conformality breaking parameter \cite{Dusling:2011fd,Romatschke:2011qp}. That said, a fully consistent formulation of hydrodynamics capturing all subtleties of the mean field contribution to its equations was provided only recently in Ref.~\cite{Czajka:2017wdo}. The analysis concerns one-component systems obeying either the classical Boltzmann statistics or the Bose-Einstein one and comprises two steps. First, the viscous correction is derived within the effective kinetic theory and then the fluid dynamics equations are obtained.

The starting point for $\Delta f$ extraction is the Boltzmann equation of the form 
\begin{eqnarray}
\label{boltz}
(\tilde k^\mu \partial_\mu -\mathcal{E}_k \nabla \mathcal{E}_k \cdot \nabla_k)
f=C[f],
\end{eqnarray}
where $C[f]$ is the collision term and $\tilde k^\mu=(\mathcal{E}_k, {\bf k})$ is the four-momentum with 
$\mathcal{E}_k = \sqrt{{\bf k}^2+\tilde m_x^2}$ and $\tilde m_x^2 \equiv
\tilde m^2(x)=m_0^2+m^2_\text{th}(x) \equiv m_0^2+m^2_\text{eq}(x)+\Delta m^2_\text{th}(x)$. The distribution function of quasiparticles is $f=f(x,k)=f_0(x,k)+\Delta f(x,k)$, where $f_0$ is of the Bose-Einstein form $f_0=(e^{E_k \beta}-1)^{-1}$ and $\Delta f$ is the nonequilibium correction which is governed by both the hydrodynamic forces and the nonequilibrium thermal mass. The correction was derived self-consistently in \cite{Czajka:2017wdo} and has the form
\begin{equation}
\Delta f = \delta f - T^2 \frac{d m^2_{\text{eq}}}{dT^2}
\frac{f_0(1+f_0)}{E_k}  
\frac{\int dK \delta f}{\int dK E_k f_0(1+f_0)},
\end{equation}
where $\delta f$ is the standard correction caused by the hydrodynamic forces, $dK\equiv d^3 {\bf k}/[(2\pi)^3 E_k]$, and $T$ is the temperature. The temperature dependence of the thermal mass is dictated by
\begin{eqnarray}
\label{mass-temp}
T^2\frac{dm^2_{\text{eq}}}{dT^2} = m^2_{\text{eq}} + a T^2 \beta_\lambda.
\end{eqnarray}
where $m^2_{\text{eq}} = \lambda T^2/24$, with $\lambda$ being the coupling. The running of the coupling is incorporated through $\beta_\lambda \equiv \beta(\lambda) = T \frac{d\lambda}{dT}$, where $a$ depends on the system consideration and for the quantum case it is $a=1/48$.

The out-of-equilibrium system is then characterized by the following stress-energy tensor
\begin{eqnarray}
\label{T-noneq}
T^{\mu\nu} &=& \int d\mathcal{K} \tilde k^\mu \tilde k^\nu f -g^{\mu\nu}U,
\end{eqnarray}
where all quantities are the nonequlibrium quantities and, in particular $d\mathcal{K}\equiv d^3{\bf k}/[(2\pi)^3 \mathcal{E}_k]$. The extra $U\equiv U(x)$-term is the nonequilibrium mean field contribution. To comply with the quasiparticle description, which assumes that the system is sufficiently dilute and the mean free path of a quasiparticle is much longer than its thermal width, the departure from the equilibrium is characterized by small corrections of all quantities from their equilibrium forms. Therefore, the function $U$ follows the expansion $U=U_0 + \Delta U$. $U_0$ is the equilibrium form of the function, which has to satisfy $dU_0= q_0 dm^2_{\rm eq}/2$ (with $q_0$ being the scalar quantity defined by $q_0=\int dK f_0$) to guarantee the thermodynamic consistency of the fluid equations and the energy and momentum conservation. Similarly, the correction $\Delta U$ has to satisfy the condition $\Delta U = q_0 \Delta m^2_{\rm th}/2$. Given that, the stress-energy tensor can be expanded in the same way, that is, $T^{\mu\nu}=T_0^{\mu\nu} + \Delta T^{\mu\nu}$ and its form given by Eq.~(\ref{T-noneq}) can be then  written as
\begin{eqnarray}
\label{T-noneq-1}
T^{\mu\nu} &=&\int dK k^\mu k^\nu f_0 - g^{\mu\nu}U_0 
+ \int dK \bigg[ k^\mu k^\nu -u^\mu u^\nu T^2\frac{dm^2_{\text{eq}}}{dT^2}
\bigg] \delta f,
\end{eqnarray}
where $u^\mu$ is the 4-velocity. The first two terms in Eq.~(\ref{T-noneq-1}) stand for the equilibrium form of the stress-energy tensor $T_0^{\mu\nu}$ and the last term is the correction. The correction determines whole dynamical information on how the equilibration process proceeds. In particular, it provides the Landau matching conditions
\begin{eqnarray}
\int dK \bigg[ (u_\mu k^\mu) k^\nu - u^\nu
T^2\frac{dm^2_{\text{eq}}}{dT^2} \bigg]\delta f =0
\end{eqnarray}
and both the shear and bulk viscous corrections
\begin{eqnarray}
\label{viscous-corr}
\pi^{\mu\nu} &=& \big\langle  k^{\langle \mu} k^{\nu \rangle}
\big\rangle_\delta , \\
\label{viscous-pi}
\Pi  &=&  -\frac{1}{3}\big\langle  \Delta_{\mu\nu} k^\mu k^\nu
\big\rangle_\delta.
\end{eqnarray}
In formulas (\ref{viscous-corr}) and (\ref{viscous-pi}) the following notation is kept: $\langle \dots \rangle_\delta \equiv \int dK (\dots) \delta f$ and
$A^{\langle \mu\nu \rangle} \equiv \Delta^{\mu\nu}_{\alpha \beta}
A^{\alpha\beta}$, where 
$\Delta^{\mu\nu}_{\alpha \beta} \equiv (\Delta^\mu_\alpha \Delta^\nu_\beta
+ \Delta^\mu_\beta \Delta^\nu_\alpha - 2/3
\Delta^{\mu\nu}\Delta_{\alpha\beta})/2$.
The definitions (\ref{viscous-corr}) and (\ref{viscous-pi}) have
well-known forms of the shear tensor and the bulk pressure, respectively, but the quantities entering them contain the thermal mass which is now position dependent. Moreover, the Landau matching condition contains a correction due to the temperature-dependent mass. 

\section{Bulk viscosity and its constraints}

The physics of bulk viscosity within different energy scales has been explored for years. However, due to the strong coupling of the matter produced in heavy ion collisions and lack of systematic methods to investigate transport phenomena at the relevant energy scale, the value of bulk viscosity of quark-gluon plasma is not easy to extract. Nevertheless, efforts to constrain its behavior are in progress. As already discussed, the expectation that bulk viscosity is related to the trace anomaly obtained from lattice QCD calculations, the QCD sum rules and the ansatz for the small frequency spectral function enabled to constrain its possible form for a deconfined nuclear matter \cite{Karsch:2007jc,Moore:2008ws,Romatschke:2009ng}. Bulk viscosity is then expected to be peaked at the temperature near the phase transition. This expectation is also confirmed by the direct lattice QCD simulations \cite{Meyer:2007dy,Meyer:2010ii,Astrakhantsev:2018oue}, which also suggest that bulk viscosity value may vary in the range $O(10^{-2})-O(10^{-1})$. The increased value of the bulk viscosity near the critical temperature was also observed exploring the low energy region of the hadron resonance gas \cite{NoronhaHostler:2008ju} and of the pion gas \cite{FernandezFraile:2008vu}.

Said that, the behavior of bulk viscosity can be studied in some detail in the extreme limits, where the coupling constant is very small or very large. A good knowledge on the bulk viscosity behavior in the extreme limits is also useful to test various analytical tools and provide benchmarks for data-driven techniques.

The high-energy limit of bulk viscosity within quantum-field theory studies was determined first for scalar theories in \cite{Jeon:1994if}, where the equivalence between the fundamental approach to transport coefficient calculations based on Kubo formulas and effective kinetic theory computations based on the Boltzmann equation was shown. The bulk viscosity was then computed and justified in terms of microscopic dynamics. In particular, it was found that the coefficient is governed by the number-changing processes. The approach based on the effective kinetic theory was then generalized to gauge theories and, in particular, the bulk viscosity of QCD matter was obtained in \cite{Arnold:2006fz}. As explicitly shown in \cite{Jeon:1994if,Arnold:2006fz} microscopic parameters which control the bulk viscosity and, at the same time, are responsible for the conformal symmetry breaking are the masses of the system's constituents and/or the Callan-Symanzik $\beta_\lambda$ function. Both parameters enter the definition of the speed of sound $c_s$, which, as a macroscopic quantity, is more conveniently used in phenomenological applications. A parametric behavior of the bulk viscosity as a function of the conformality breaking parameter is then expressed as
$\frac{\zeta}{\eta} \propto \Big(\frac{1}{3} - c_s^2 \Big)^2$,
where the shear viscosity $\eta \propto T^3/(\alpha_s^2 \log(1/\alpha_s))$, see Ref.~\cite{Arnold:2006fz}.

On the other hand, approaches based on holography, applicable to study strongly coupled systems, where the coupling constant goes to infinity, provide a linear dependence of the ratio $\zeta/\eta$ on the conformal symmetry breaking parameter, $\frac{\zeta}{\eta} \propto \Big(\frac{1}{3} - c_s^2 \Big)$, with $\eta/s=1/(4\pi)$, see Ref.~\cite{Benincasa:2005iv}. Importantly, in Ref.~\cite{Buchel:2007mf} the lower bound on bulk viscosity was obtained to be $\zeta/\eta \geq 2(1/3-c_s^2)$.

Recent progress on bulk viscosity studies has had many aspects. In Ref.~\cite{Czajka:2018bod} bulk viscosity was studied in the whole range of the 't~Hooft coupling. In particular, the justification of the effective kinetic theory for bulk viscosity computation was done in the large $N_c$ limit. A diagrammatic analysis and comprehensive power counting of processes contributing to the collision kernel of the Boltzmann equation were provided. Both elastic and inelastic processes were considered and the corresponding integral equations were schematically advocated. In the strong coupling side the holographic methods from string theory duals and M-theory were applied to investigate the bulk viscosity physics for the weak and strong string couplings, respectively. In both cases, the ratio $\zeta/\eta$ is found to be linearly proportional to the deviation from conformality and different microscopic dynamics in each case results in different numerical factors, but both satisfy Buchel bound \cite{Buchel:2007mf}.

In Refs.~\cite{Denicol:2014vaa,Florkowski:2015lra} the bulk viscosity behavior was studied for one-component gases of different statistics in the relaxation time approximation when the conformal anomaly is caused only by the constant mass of particles composing a system. These results were later updated in Ref.~\cite{Czajka:2017wdo}, where the mean-field effects are systematically incorporated and consequences of the quantum phase space density are discussed in some detail. Two common frameworks of kinetic theory were applied to obtain the coefficients emerging in the bulk mode. First, Anderson-Witting model, that is, the relaxation time approximation of Chapman-Enskog method, was used to find the ratio of the bulk viscosity to the relaxation time $\zeta/\tau_R$. Then, the same ratio and other transport coefficients arising in the equation of motion for bulk pressure were studied within 14-moment approximation. Both methods provide the same results on $\zeta/\tau_R$. For the Boltzmann statistics the ratio $\zeta/\tau_R$ reflects the expected parametric form of the source of the scale symmetry breaking $\zeta_{\rm Boltz}/ \tau_R \propto T^4 (1/3 - c_s^2)^2$. It is also explicitly shown that the conformality breaking parameter $(1/3-c_s^2)$ is directly related to the combination of the microscopic parameters responsible for the conformal anomaly: the constant mass of the system's constituents and the $\beta_\lambda$ function. For a system with Bose-Einstein distribution function, the result is
\begin{equation}
{\zeta \over\tau_R}
\propto
T^4 \left({1\over 3} - c_s^2\right)^2 {T \over m_{x}},
\end{equation}
where $m_x$ includes both the zero-temperature mass and the thermal mass.
The ratio differs from the expected form by the factor $T/m_x$, which is a consequence of the cut-off of infrared divergences. Such a dependence is attributed to the application of the relaxation time approximation, which seems to be too crude to compute the ratio reliably. Other coefficients appearing in the bulk mode feature even stronger dependence on the factor $T/m_x$.

Consequences of nonconformality were also studied within the hydro-kinetic theory in Ref.~\cite{Akamatsu:2017rdu}, where the effect of hydrodynamic fluctuations was taken into account. The fluctuations occurring out of equilibrium are shown to modify the stress-energy tensor. It then leads to renormalization of hydrodynamic forces as well as the bulk viscosity. The renormalization of bulk viscosity is shown to be proportional to the scale symmetry breaking in the equation of state and is found to be 
\begin{equation}
\zeta(T)= \zeta_0(T,\Lambda)+\frac{T\Lambda }{18\pi^2} 
\Big[ \Big(1+ \frac{3T}{2} \frac{dc^2_{s0}}{dT} -3c^2_{s0} \Big)^2 \frac{e_0+p_0}{\zeta_0+\frac{4}{3}\eta_0} + 4(1-3c^2_{s0})^2 \frac{e_0+p_0}{2\eta_0} \Big],
\end{equation}
where $\zeta_0(T,\Lambda)$ is the bare bulk viscosity and $\Lambda$ is the ultraviolet cut-off of hydrodynamic fluctuations. The results are claimed to be helpful in constraining minimal value of bulk viscosity of hot QCD.

\section{Summary}

This review summarizes recent progress in the description of strongly interacting matter, fluid dynamics development and physics of conformal invariance breaking. Conformal anomaly and finite bulk viscosity have a significant effect on hydrodynamic modeling of strongly interacting matter produced in nuclear collisions. The sensitivity of many physical observables to a size and temperature dependence of the bulk viscosity has recently been tested within different models. Besides the impact of bulk viscosity on the quark-gluon plasma dynamics, the physics of various nonconformal systems at different energy scales is also reviewed. Such studies largely contribute to testing different approaches and provide constraints on the bulk viscosity and physics governing its behavior.

\end{document}